\documentclass[a4paper,fleqn,usenatbib]{mnras}
\usepackage{newtxtext,newtxmath}
\usepackage[T1]{fontenc}
\usepackage{ae,aecompl}
\usepackage{graphics}
\usepackage{graphicx}
\usepackage{amsmath}	
\usepackage[flushleft]{threeparttable}
\usepackage{booktabs,caption}
\usepackage{placeins}
\usepackage{float}

\title[SXP 15.3]{ Super-critical accretion of BeXRB SXP 15.3 }

\author[Ghising et. al.]{
Manoj Ghising,$^{1}$\thanks{manojghising26@gmail.com}
Binay Rai,$^{1}$\thanks{binayrai21@gmail.com}
Mohammed Tobrej,$^{1}$\thanks{tabrez.md565@gmail.com} 
Ruchi Tamang,$^{1}$\thanks{ruchitamang76@gmail.com}
\newauthor
Bikash Chandra Paul$^{1}$\thanks{bcpaul@nbu.ac.in}
\\
$^{1}$Department of Physics, North Bengal University,Siliguri, Darjeeling, WB, 734013, India
\\
}
\pubyear{2021}

\begin{document}
\label{firstpage}
\pagerange{\pageref{firstpage}--\pageref{lastpage}}
\maketitle

\begin{abstract}
We have studied the temporal and spectral properties of SXP 15.3 observed by \textit{NuSTAR} in hard energy range 3-79 keV during late 2018. The timing analysis of \textit{NuSTAR} observation predicts coherent pulsation at $15.2388\;\pm\;0.0002\;s$. The pulse profiles in different energy bands demonstrates energy dependence. The shape of the pulse profile was generally suggestive of a fan-beamed dominated pattern, which when combined with the measured luminosity predicts that the source may be accreting in the super-critical regime. Non-monotonic increase of Pulse Fraction was observed with energy. The \textit{NuSTAR} observation finds that the pulse period of the source have spun-up at a rate of --0.0176 s $yr^{-1}$ when compared to the previous analysis by the same observatory more than 1 year ago. The source flux in the present \textit{NuSTAR} study in 3-79 keV energy range is $\sim\;1.36\;\times\;10^{-10}\;erg\;cm^{-2}\;s^{-1}$ which corresponds to a luminosity of $\sim\;6\;\times\;10^{37}\;erg\;s^{-1}$. The cyclotron line energy of the source is detected at $\sim$5 keV. Pulse phase resolved spectroscopy shows that the cyclotron line energy varies significantly with pulse phase and the photon index becomes softer with increasing flux. In addition, we have studied the evolution of luminosity with time using 2017 and 2018 \textit{Swift/XRT} observations. The analysis of \textit{Swift/XRT} data reveals that the photon index are positively correlated with the source luminosity which is a characteristic of super-critical accretion phenomena.  

\end{abstract}

%% Select between one and six entries from the list of approved keywords.
%% Don't make up new ones.
\begin{keywords}
accretion, accretion discs-stars: neutron-pulsars: individual: SXP15.3: individual: X-rays: binaries .
\end{keywords}

%%%%%%%%%%%%%%%%%%%%%%%%%%%%%%%%%%%%%%%%%%%%%%%%%%%
%
%%%%%%%%%%%%%%%%%% BODY OF PAPER %%%%%%%%%%%%%%%%%%
%
\section{Introduction}

Based on the mass of the companion star, they are classified into two categories viz., High Mass X-ray Binaries (HMXBs) and Low Mass X-ray Binaries (LMXBs). HMXBs are further classified into two categories \textit{viz.} Be/X-ray Binaries (BeXRBs) and Supergiant X-ray Binaries \citep{Reig2011}. High-mass X-ray binaries (HMXB) are binary systems that constitute a compact object, generally a neutron star (NS) along with a high mass optical companion. Be/X-ray binaries (BeXRBs; see review - \citep{Reig2011}) are a major subclass of HMXBs, constituting the majority of the known accreting X-ray pulsars with a magnetic field strength of the order $10^{12}$ G or above \citep{Ikhsanov_Mereghetti2015}. 

The Small Magellanic Cloud (SMC) is known to contain a large collection of BeXRBs \citep{Coe_Kirk2015,Haberl_and_Sturm2016}. A low Galactic foreground absorption and a well-measured distance of $\sim$62 kpc \citep{Graczyk2014} provides ample opportunity for exploring X-ray binary systems in the SMC. The SMCs  are known for hosting a large number of HMXBs as compared to Large Magellanic Cloud (LMC) \citep{Haberl_Sasaki2000,Yokogawa2003}. 
%The catalog survey of \cite{Haberl_Pietsch2004} shows 65 HMXBs and candidates in the SMC, out of which at least 37 HMXBs show X-ray pulsations, indicating the spin period of the neutron star. 
The BeXRB system is composed of a neutron star and a fast-rotating early-type star with an equatorial circumstellar disc \citep{Reig2011}. The optical counterpart of a BeXRB system is a \textit{Be} star. The \textit{Be} star is a non-supergiant fast-rotating B-type which belongs to luminosity class III-IV stars. The letter "\textit{e}" associated with "\textit{Be}" is due to spectral line emission which is primarily the Balmer series \citep{Porter_Rivinius2003}. \textit{Be} stars can also exhibit He and Fe lines in addition to their defining Hydrogen lines \citep{Hanuschik1996}. The rapid rotation of a companion star \textit{Be} in a BeXRB system leads to the expulsion of photospheric matter along its equatorial plane. As a result, a circumstellar disk is formed around it. The circumstellar disk evolves continuously \citep{Okazaki2001} thereby allowing the neutron star to capture matter at certain times. When the Neutron Star (NS) makes a periastron passage \textit{i.e} closest approach to the \textit{Be} star, the accretion activity is enhanced significantly thereby resulting in enhanced X-ray emission. However, the X-ray emission activity drops by several magnitudes when a NS moves away from the periastron. This implies that the system tends to approach the quiescent state. These systems exhibit transient character and are best studied when the source undergoes bright outbursts.  
The correlation between the orbital period and spin period of the neutron star is one of the characteristics of these systems \citep{Corbet1984}. In general, the spin period distribution of these binary system follows a bimodal distribution. The origin of such distribution has been proposed based on two different types of supernovae \textit{viz.} iron-core-collapse and electron-capture \citep{Knigge_Coe_Podsiadlowski2011}. Previous studies by \cite{Cheng_Shao_Li_2014} have proposed other possible linkages between a  bimodal distribution and different accretion regimes. According to \cite{Cheng_Shao_Li_2014}, when the NS undergoes giant outbursts and tends to absorb material from the deformed, outer region of the Be star disc, a disc dominated by radiative cooling forms around the NS, spinning it up and giving rise to the short-period subpopulation. The accretion flow is advection-dominated or quasi-spherical in BeXRBs that are dominated by regular or persistent outbursts. As a result, the spin-up mechanism is ineffective, resulting in longer periods of the neutron stars. However, the later proposed schemes have been considered and supported by different long-term X-ray variability of the binary system with both short and long orbital periods \citep{Haberl_and_Sturm2016}.

The X-ray source SXP 15.3 (aka RX J0052.1-7319) located in the SMC was first detected with observatory Einstein (\cite{WangWu1992}). Its transient character was classified by \cite{Kahabka_Pietsch1996} based on the \textit{ROSAT PSPC} data. The optical counterpart of the source is a B-type star which shows $H_{\alpha}$ emission  \citep{Israel1999}. It was later confirmed as an O9.5 IIIe star \citep{Covine2001}. The long-term optical variability of the source was reported by \cite{Udalski1999} based on Optical Gravitational Lensing Experiment (OGLE) monitoring. The X-ray pulsations for this source was discovered during \textit{ROSAT HRI} and \textit{BATSE} observations \citep{Lamb1999}. The source outburst was not reported before 2017, when the Swift SMC Survey concluded a brightening of the source \citep{Kennea2017}. Detailed work on this source during the bright outburst was performed by \cite{Maitra2018} where a 5 keV absorption feature known as Cyclotron Resonance Scattering Feature (CRSF) was reported. 

The CRSF feature originates from a region close to the surface of a neutron star, particularly at the poles where the magnetic field strength is strong (for reviews on CRSF, see \cite{Staubert2018,Maitra2018}). When the matter accreted by the NS falls on the magnetic poles, it results in the conversion of the kinetic energy of the infalling matter into heat and radiation. Moreover, the electrons present in the infalling matter in a direction perpendicular to the magnetic field gets quantized in discrete energy levels referred as Landau levels \citep{Schonherr2007}. Those photons whose energy lies close to the Landau levels gets scattered by the quantized electrons and as a result, a cyclotron line is formed in the spectra. The estimated centroid energy of the cyclotron line can be used for direct estimation of the magnetic field of a neutron star using the relation, $B\;=\;\frac{E_{c}\;(1+z)}{11.57}10^{12} \;G$, where $E_{c}$ is the centroid energy of the CRSF in units of keV \citep{Harding_Daugherty1991} and \textit{z} is the Gravitational redshift parameter.
  The range of energy of the cyclotron lines can vary from as low as $\sim$5 keV for the source SXP 15.3 \citep{Maitra2018} to about 90 keV for the sources LMC X-4 \citep{11} , and GRO J1008-57 \citep{chen2021}. 
 In some X-ray pulsars,  it is found that higher harmonics  present in addition to the fundamental cyclotron line \citep{13,14,15,16}. The cyclotron line energy varies with the pulse phase which can be confirmed from the phase-resolved spectroscopy. It is interesting to note that in recent analysis  of the X-ray pulsars GRO J2058+42 \citep{21}; Swift J1808.4-1754 \citep{22}), the CRSF feature although absent in the average continuum spectra, can be present only at certain phases.

 The motivation of this paper is to probe the detailed broad-band coverage of  temporal and spectral analysis of SXP 15.3 in late 2018 using \textit{NuSTAR} observations. In addition, for a comparative study of the source, we have also considered \textit{NuSTAR} observation corresponding to the outburst in 2017 as well as \textit{Swift} data. In section 2, we have explained the data reduction formalism corresponding to \textit{NuSTAR} observation. The results of timing and spectral analysis of the source are explained in detail in section 3 and section 4. Discussions and conclusions have been presented in section 5. 
 
\section{Observations and Data Reduction}

\subsection{\textit{NuSTAR}}
The data reduction for SXP 15.3 has been carried out using \texttt{heasoft}\footnote{\url{https://heasarc.gsfc.nasa.gov/docs/software/heasoft/download.html}} v6.30.1 and \textsc{caldb} v20220803. \textit{NuSTAR} was the first hard X-ray focusing telescope operating in the energy range of (3-79) keV. It consists of two identical X-ray telescope modules that have their own focal plane modules A \& B  (\texttt{FPMA and FPMB}) consisting of a  solid-state CdZnTe detector (\cite{Harrison2013}).
The extraction \& screening of data has been carried out using software \texttt{nustardas}  v2.1.1. The unfiltered clean event files were filtered using the  mission-specific command \texttt{nupipeline}. Using \texttt{xselect} command, we extracted the clean event files of the respective instrument \texttt{FPMA and FPMB}. The image was then extracted with the help of \texttt{Ds9}\footnote{\url{https://sites.google.com/cfa.harvard.edu/saoimageds9}} application software. A circular region of  49" around the source center and the background region of the same size away from the source region were selected from chip numbered 0 and 1 as the source and background region files respectively. Next, we run \texttt{nuproducts} using the source and background region files to get the necessary light curves \& spectra. Background subtracted light curves were then obtained using \textit{ftool} \texttt{lcmath} for both the instrument \texttt{FPMA and FPMB}. Barycentric corrections were then performed using \textit{ftool} \texttt{barycorr}. We made use of \textit{NuSTAR} observation (see Table 1) for performing the required timing and spectral analysis of the system.

\subsection{\textit{Swift}}
The \textit{Swift} observatory \citep{37} carries three scientific instruments viz. Burst Alert Telescope(BAT), X-ray Telescope (XRT) and UV/Optical Telescope (UVOT), not considered in this analysis). The three instruments onboard \textit{Swift} covers a broad energy range of $\sim$ 0.002-150 keV. \textit{Swift} (15-50 keV) performs regular monitoring of the X-ray sky \citep{31}. \textit{Swift}/XRT (0.5-10 keV) performed pointed observations of SXP 15.3 during the outburst episode, starting at 57960 MJD.
\textit{Swift}/XRT data were reduced using mission-specific task \texttt{xrtpipeline} provided by the \texttt{heasoft} v6.30.1 and using the \texttt{caldb} v20210915. Filtering and screening criteria were done using \textit{ftools}. 
. Using \textit{ftool} \texttt{xselect}, we extracted event files and data counts for the respective  observation IDs. A source region of 30" and background region of the same size for observations in Window Timing Mode while an annulus region with outer radius of 20" and inner radius of 10" for observations in Photon Counting Mode  were considered using astronomical imaging and data visualization application \texttt{ds9}\footnote{\url{https://sites.google.com/cfa.harvard.edu/saoimageds9}} for extracting necessary light curves and spectra. For spectral analysis, we generated \texttt{ancillary response files (arf)}  with \texttt{xrtmkarf} to account for different extractions region, \texttt{PSF} corrections and vignetting. The source spectra  was analyzed in \texttt{xspec} v12.12.1 \citep{45}.

\begin{table}
 \begin{center}
 \begin{tabular}{clllc}
    \hline
    
   Observatory	& Date of observation &	OB IDs	&	Exposure	\\
	&		&	& (in ksec)	\\
\hline	
			& 2017-11-30 & 30361002002 &	70.65 \\
\textit{NuSTAR}	   &  &  & \\
 			&	2018-10-27 & 30361002004	&	58.35	\\

      \hline
  \end{tabular}
  \caption{\textit{NuSTAR} observation details of the source SXP 15.3.}
  \end{center}
  \label{13}
 \end{table}

\section{TIMING ANALYSIS}

\begin{figure}

\begin{center}
\includegraphics[angle=270,scale=0.3]{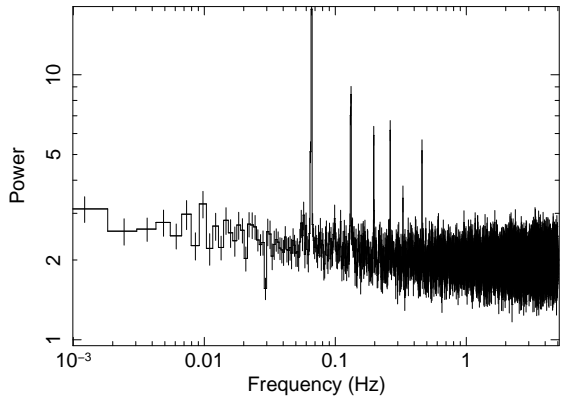}
\end{center}
\caption{FFT of the \textit{NuSTAR} 3-79 keV energy range light curve  of the source SXP 15.3 revealing the presence of a harmonic along with the fundamental signal.}
\label{1}
\end{figure}

\subsection{Pulse period estimation}

\begin{figure}

\begin{center}
\includegraphics[angle=0,scale=0.3]{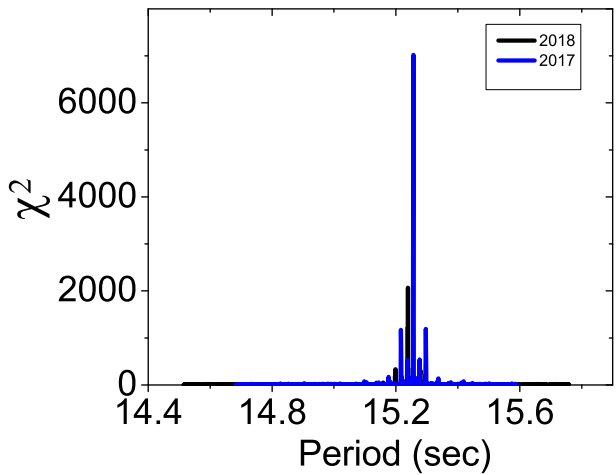}
\end{center}
\caption{Periodogram of the source SXP 15.3 corresponding to 2017 (blue) and
2018 (black) \textit{NuSTAR} observation in the 3-79 keV energy range. A coherent signal is clearly detected
at $\sim$ 15.2388 s and $\sim$ 15.2564 s. }
\label{2}
\end{figure}

For timing analysis, we consider barycentric-corrected \textit{NuSTAR} light curves of the source SXP 15.3 with binning 0.01 s. In order to obtain the approximate pulse period of the source, we made use of Fast Fourier Transform (FFT) using \textit{ftool} \texttt{powspec} of the light curve as shown in Figure \ref{1}. The FFT of the light curve reveals the presence of a significant harmonic peak at half the fundamental pulse period. The pulse period was determined by the epoch-folding technique \citep{Davies1990,Larsson1996} using ftool \texttt{efsearch} while taking into account the rough pulsation that had previously been obtained by FFT. This technique is based on $\chi^{2}$ maximization technique and hence we estimated the precise period at 15.2388\;$\pm$\;0.0002 s (see Figure \ref{2} marked in black color). The uncertainty in the measurements was obtained by employing the method described by \cite{Boldin2013}. For this, we generated 1000 simulated light curves and obtained the pulse period of each curve by using \textit{ftool} \texttt{efsearch}. Finally, we estimated the mean and the standard deviation of 1000 pulsations to arrive at  15.2388\;$\pm$\;0.0002 s. In order to understand the spin-period evolution of the source, we took the 2017 outburst data. By analyzing in the same way as described above, we arrived at 15.2534 $\pm$ 0.0004 s. The two observations being almost one year apart, the pulsar has spun up by --0.0176$\pm0.0002$ s $yr^{-1}$.

\subsection{Pulse Profile} 
\begin{figure}

\begin{center}
\includegraphics[angle=0,scale=0.3]{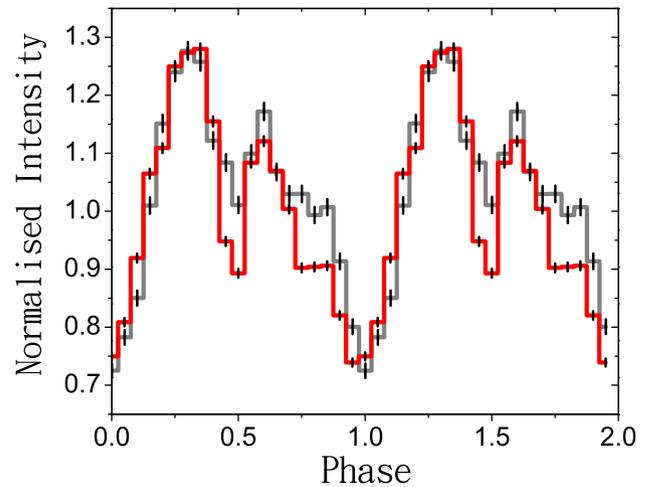}
\end{center}
\caption{Folded pulse profile of SXP 15.3 in the energy range (3-79) keV. The red color corresponds to 2017 outbursts while grey corresponds to 2018 \textit{NuSTAR} data. }
\label{3}
\end{figure}

\begin{figure*}

\centering
\includegraphics[angle=0,scale=0.5]{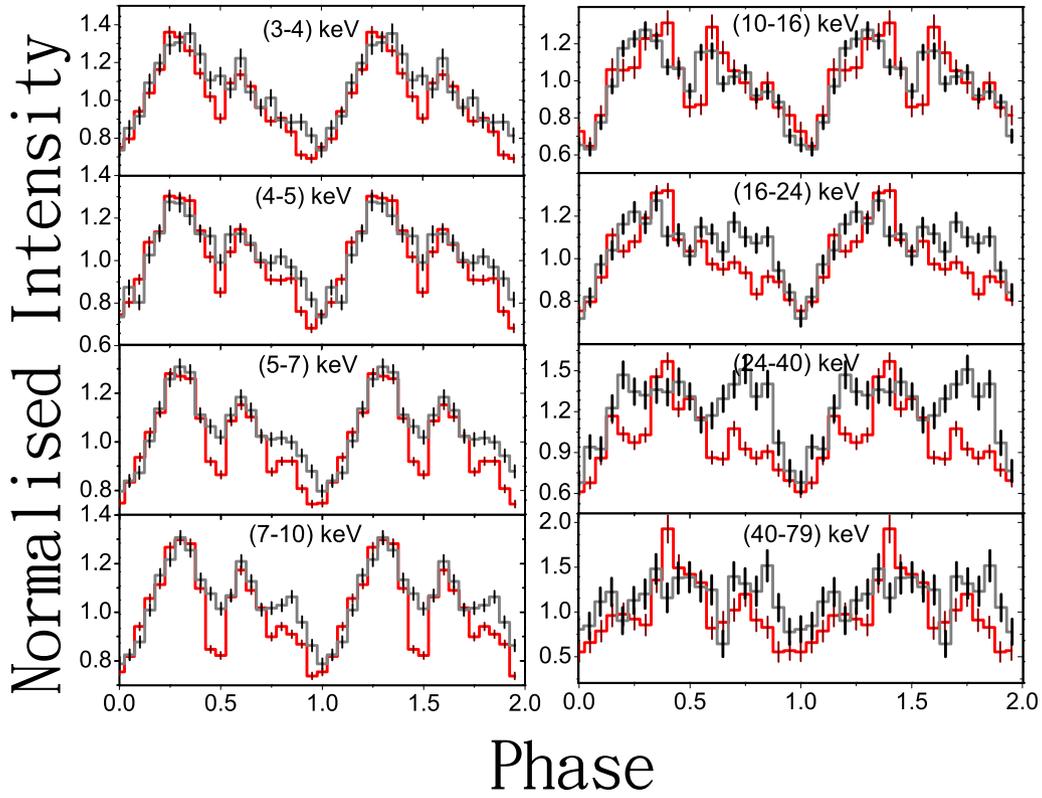}

\caption{Folded pulse profile of SXP 15.3 in several energy bands. The red color corresponds to 2017 outbursts while grey corresponds to 2018 \textit{NuSTAR} data. }
\label{4}
\end{figure*}

\begin{figure}

\begin{center}
\includegraphics[angle=0,scale=0.3]{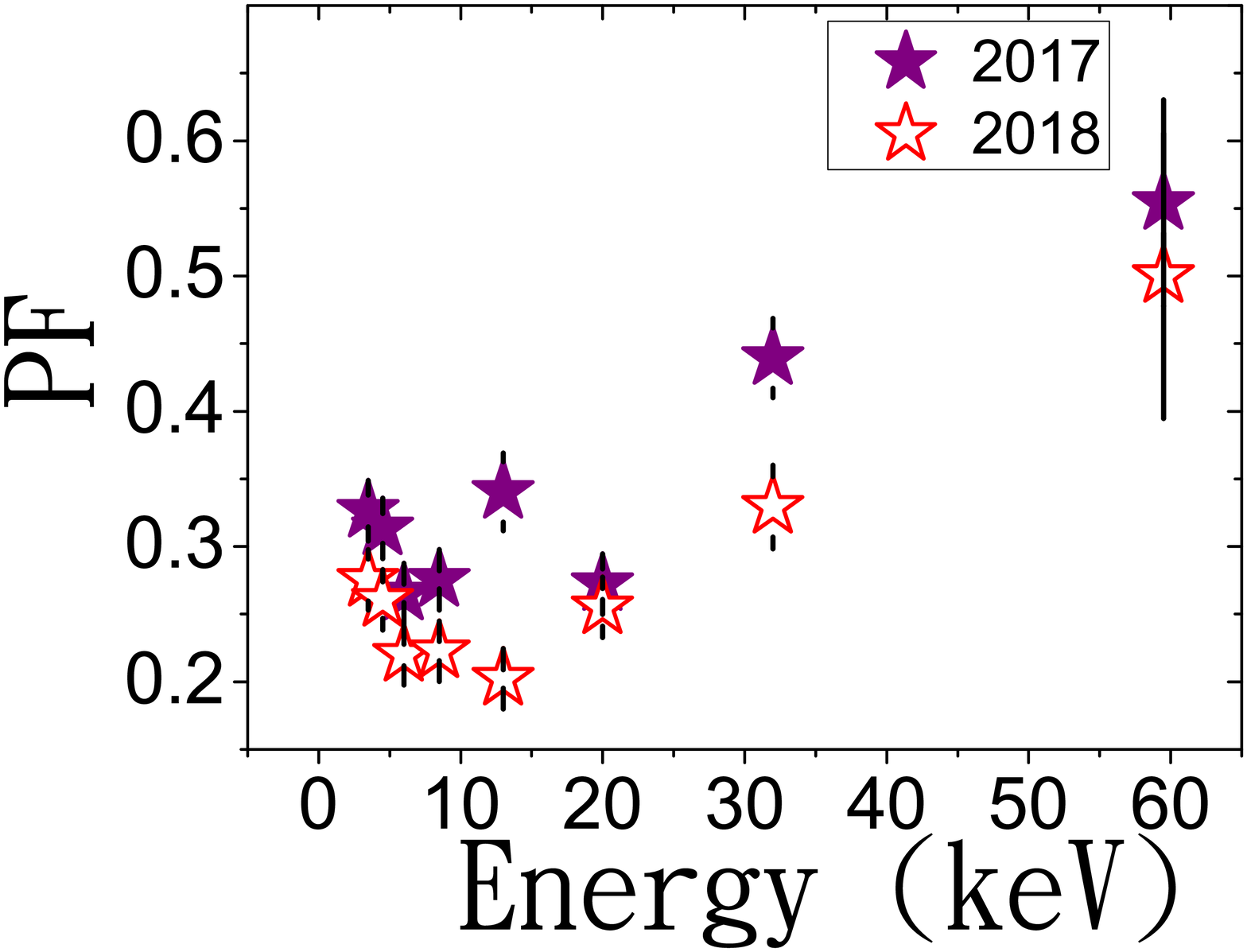}
\end{center}
\caption{Variation of Pulse Fraction (PF) with energy of SXP 15.3.}
\label{5}
\end{figure}

\begin{figure}

\begin{center}
\includegraphics[angle=0,scale=0.3]{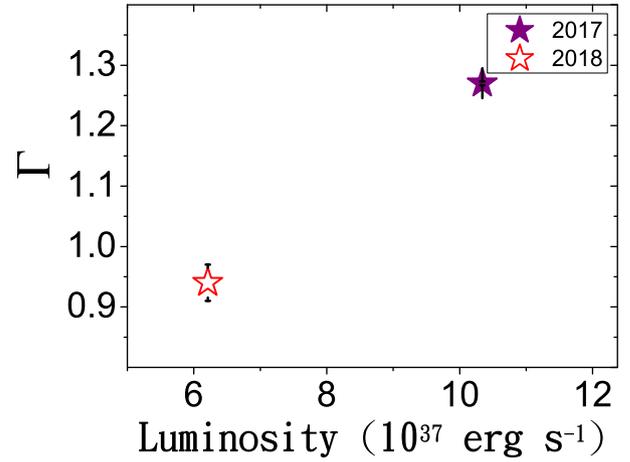}
\end{center}
\caption{Variation of Photon Index ($\Gamma$) with luminosity in the energy range (3-79) keV. }
\label{6}
\end{figure}

In order to study the variation of Pulse Profile with energy, we resolved the light curve in the energy range (3-79) keV into several energy bands of (3-4) keV, (4-5) keV, (5-7) keV, (7-10) keV, (10-16) keV, (16-24) keV, (24-40) keV and (40-79) keV. We have arbitrarily chosen the zero-point of the profiles such that the minimum bin lies at phase=0.0. The \textit{NuSTAR} pulse-profile corresponding to 2017 and 2018 in the broadband energy range (3-79) keV is shown in Figure \ref{3}. The pulse profiles in different energy bands as shown in Figure \ref{4} demonstrate clear dependence on energy. Initially, the pulse profile at lower energy band (3-4) keV remains single-peaked and then evolves weakly into double-peaked in the lower intermediate bands (4-5) keV and (5-7) keV and then loses its double peak shape above 7 keV. The main peak emission lies close to phase 0.3 while the additional emission component beginning at phase 0.6 in the lowest energy band (3-4) keV becomes broader with energy.

In order to understand the  evolution of pulse profile with time, we compared it with observations of the 2017 outburst. It is obvious from Figure \ref{4}, that the pulse profile has evolved with time. At low energies,  the morphology of the pulse profile is transformed from double-peaked nature in 2017 to a single  peaked nature in 2018. A transition from double peak to single was observed in the 2017 outburst, however, in the one year interval, the profile undergoes a transition from the single peak in the lowest band to double peak in the intermediate bands and then again makes the transition back to a broad single peak at higher energy bands. However, the transition from double peak to single peak occurs at earlier energy bands than the transition that occurred a year ago.

\subsection{Pulse Fraction}
The Pulse Fraction (PF) is defined as the ratio of the difference between the maximum and minimum intensity (\texttt{$C_{MAX}-C_{MIN}$}) to the sum of the maximum and minimum intensity ($C_{MAX}+C_{MIN}$) of the pulse profile i.e $PF\;=(C_{MAX}-C_{MIN})/(C_{MAX}+C_{MIN})$. The PF initially shows a decreasing trend with energy up to $\sim$ 14 keV and after that, it increases monotonically with energy. Prominent local minima cannot be observed in the present work near (5-8) keV and 20 keV as was observed in 2017 outbursts (see Figure \ref{5}). The maximum PF between 2017 and 2018 \textit{NuSTAR} data was 0.55$\pm$0.07 and 0.50$\pm$0.10 while minimum PF was 0.26$\pm$0.01 and 0.20$\pm$0.01. It can be seen from Figure \ref{6} that the PF has increased with the source luminosity.

\section{Spectral Analysis}

\begin{table*}
\begin{center}
\begin{tabular}{clllllllllllc}
\hline	

Parameters	&	&	&	\textsc{model i} (cutoffpl)	&	&	&	\textsc{model ii} (highecut)	&	&	&	\textsc{model iii} (compTT)	\\
\hline													
$C_{FPMA}$	&	&	&	1(fixed)	&	&	&	1(fixed)	&	&	&	1(fixed)	\\
$C_{FPMB}$	&	&	&	0.997$\pm$0.005	&	&	&	0.997$\pm$0.005	&	&	&	0.997$\pm$0.005	\\
$n_{H}\;(\times10^{22}\;cm^{-2})$	&	&	&	0.81 (fixed)	&	&	&	0.81 (fixed)	&	&	&	0.81 (fixed)	\\
compTT ($T_{o}$) (keV)	&	&	&	-	&	&	&	-	&	&	&	0.78$\pm$0.06	\\
compTT (kT) (keV)	&	&	&	-	&	&	&	-	&	&	&	8.10$\pm$0.20	\\
compTT ($\tau$) (keV)	&	&	&	-	&	&	&	-	&	&	&	3.98$\pm$0.10	\\
$\Gamma$	&	&	&	0.94$\pm$0.02	&	&	&	0.94$\pm$0.03	&	&	&	-	\\
$E_{CUT}$ (keV)	&	&	&	18.23$\pm$0.72	&	&	&	7.06$\pm$0.43	&	&	&	-	\\
$E_{fold}$ (keV)	&	&	&	-	&	&	&	21.95$\pm$0.66	&	&	&	-	\\
Fe line (keV)	&	&	&	6.41$\pm$0.04	&	&	&	6.40$\pm$0.06	&	&	&	6.41$\pm$0.03	\\
$\sigma_{Fe}$ (keV)	&	&	&	0.1 (fixed)	&	&	&	0.1 (fixed)	&	&	&	0.1 (fixed)	\\
$E_{CYC}$ (keV)	&	&	&	4.82$\pm$0.16	&	&	&	4.83$\pm$0.16	&	&	&	4.13$\pm$0.18	\\
$\sigma_{CYC}$ (keV)	&	&	&	2.68$\pm$0.14	&	&	&	2.67$\pm$0.14	&	&	&	2.49$\pm$0.09	\\
$Strength_{CYC}$ keV	&	&	&	1.44$\pm$0.16	&	&	&	1.34$\pm$0.14	&	&	&	2.74$\pm$0.24	\\
Flux ($\times   10^{-10}\;erg\;cm^{-2}\;s^{-1}$)	&	&	&	1.37$\pm$0.02	&	&	&	1.37$\pm$0.03	&	&	&	1.34 $\pm$0.03	\\
$\chi^{2}_{\nu}$	&	&	&	1.05	&	&	&	1.04	&	&	&	1.07	\\

 \hline
  \end{tabular}
  \caption{The above table represents the best-fit parameters of SXP 15.3 using the continuum models constant$\times$TBabs$\times$(cutoffpl+gaussian)$\times$gabs, constant$\times$TBabs$\times$(highecut$\times$powerlaw+gaussain)$\times$gabs and constant$\times$TBabs$\times$(compTT+gaussian)$\times$gabs. $n_{H}$ represents equivalent Hydrogen column density, $\Gamma$ and $E_{CUT}$ represents Photon Index  and cutoff energy of cutoffpl model, $T_{o}$, kT and $\tau$ represents soft comptonization temperature, plasma temperature and plasma optical depth of compTT model, Fe and $\sigma$ represents iron line and its equivalent width of gaussian model, $E_{fold}$ represents folded energy of highecut model and $E_{CYC}$, $\sigma_{CYC}$ and $Strength_{CYC}$ represents the Cyclotron line energy, width and the strength of the line corresponding to gabs component. Flux were calculated within energy range (3-79) keV for the \textit{NuSTAR} observation. The fit statistics $\chi_{\nu}^{2}$  represents reduced $\chi^{2}$ ($\chi^{2}$ per degrees of freedom). Errors quoted for each parameter are within 90\% confidence interval.}
  \end{center}
 \end{table*}
\subsection{Phase-average spectroscopy}

\begin{figure}

\centering
\includegraphics[angle=270,scale=0.3]{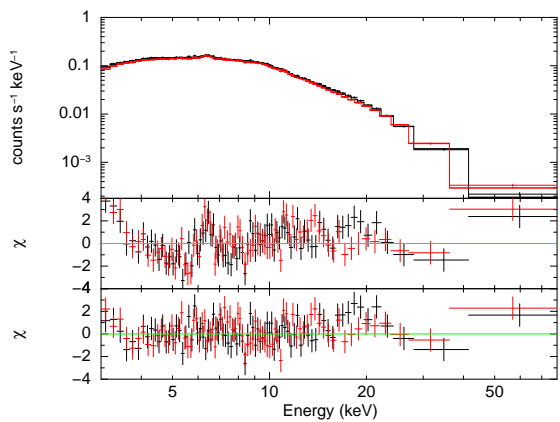}

\caption{Energy spectrum of \textit{NuSTAR} observation using the continuum model constant$\times$TBabs$\times$(cutoffpl+gaussian)$\times$gabs. The middle panel represents the residuals without gaussian and gabs component. The bottom panel represents residuals after including the gaussian and gabs component. Black \& red spectra corresponds to \textit{NuSTAR} \texttt{FPMA and FPMB} respectively.}
\label{7}
\end{figure}

Spectral analysis of the source SXP 15.3 was performed using \textsc{xspec} v12.12.1 \citep{45} in the broad energy range (3-79) keV. The \textit{NuSTAR} \texttt{FPMA and FPMB} data were grouped to have at least 20 counts per spectral bin using \textsc{xspec} tool \textsc{grppha}. The normalization factor between the two modules were ensured by introducing a model \textsc{constant} whereby the constant parameter of \texttt{FPMA} was fixed to unity while for \texttt{FPMB}, the parameter was kept free. TBabs model with abundance from \cite{Wilms2000} were introduced to take into account the interstellar absorption and gaussian model to take into account the emission line of Fe. The atomic cross-section in TBabs were adapted from \cite{Verner1996} during the spectral fit. The spectra of the source were fitted using three different model combinations constant$\times$TBabs$\times$(cutoffpl+gaussian)$\times$gabs, constant$\times$TBabs$\times$(highecut$\times$powelaw+gaussain)$\times$gabs and constant$\times$TBabs$\times$(comptt+gaussian)$\times$gabs. In order to take into   account the local absorption within the SMC, we introduced pcfabs component. However, the component parameters were inconsistent and did not improve the fit. Hence, we dropped this component. The three model combinations are referred to as \textsc{model i} , \textsc{model ii}  and \textsc{model iii} (see Table 2). Since the \textit{NuSTAR} spectra is well calibrated above 3 keV, therefore, we could not constrain the neutral hydrogen column density ($n_{H}$). We, kept $n_{H}$ fixed for all the model combinations at the Galactic value $0.81\;\times\;10^{22}\;cm^{-2}$  in the direction of the source \citep{HI4PI2006}. The three model combination fits the spectra very well with observed $\chi_{\nu}^{2}$ of 1.05, 1.04, and 1.07 respectively (see Table 2). The flux of the system is $\sim1.36\;\times\;10^{-10} \;erg\;cm^{-2}\;s^{-1}$ which corresponds to a luminosity of $\sim6\;\times\;10^{37}\;erg\;s^{-1}$ assuming a distance of 62 kpc \citep{Graczyk2014}. The broadband spectral fit in the (3-79) keV energy range corresponding to \textsc{model i} has been shown in Figure \ref{7}.

\begin{table*}
\begin{center}
\begin{tabular}{clllllllllllc}
\hline								
\hline								
OB Ids	&	& TIME (MJD)	&	& 	Exposure (sec)	& & MODE\\
\hline
\hline
00010210002	& &	58048.41		& & 0.93 & & WT	\\
00010210004	& &	58066.64	 & & 0.66 & & WT		\\
00010210005	& &	58069.63		& & 0.99 & & WT	\\
00010210007	& &	58075.72		& & 1.59 & & WT	\\
00010210008	& &	58078.04		& & 0.89 & & WT	\\
00010210009	& &	58081.03	 & & 1.33	& & WT	\\
00010210010	& &	58084.02		& & 0.78 & & WT	\\
00088639001	& &	58087.61		& & 1.54 & & PC	\\
00010210011	& &	58090.33		& & 1.23 & & WT	\\
00010210012	& &	58097.09		& & 1.27 & & WT	\\
00010210013	& &	58112.47		& & 0.63 & & WT	\\
00010210014	& &	58115.72	& & 0.88 & & WT	\\
00010210015	& &	58118.44		& & 1.29 & & WT 	\\
00010210016	& &	58123.76		& & 0.45 & & WT	\\
00010210017	& &	58130.19		& & 1.40 & & WT	\\
00010210018	& &	58134.03		& & 0.58 & & WT	\\
00010210019	& &	58136.50		& & 1.38 & & WT	\\
00010210020	& &	58142.68		& & 1.04 & & WT 	\\
00010210021	& &	58145.26		& & 0.83 & & WT	\\
00010210022	& &	58149.04		& & 0.86 & & WT	\\
00010210023	& &	58167.40		& & 0.69 & & WT	\\

00088639002	& &	58418.85		& & 0.40 & & PC	\\
00088639003	& &	58419.05		& & 0.32 & & PC	\\
00010210034	& &	58411.18		& & 1.03 & & PC	\\
00010210035	& &	58424.10		& & 0.72 & & PC	\\
00010210036	& &	58427.80		& & 0.63 & & PC	\\
00010210037	& &	58430.33	& & 0.62 & & PC	\\

\hline								
\hline

 \end{tabular}
  \caption{Observtion details of \textit{Swift/XRT} data from (57960-58419) MJD. WT represents Windowed Timing mode and PC represents Photon Counting mode.}
  \end{center}
 \end{table*}

\begin{figure}

\begin{center}
\includegraphics[angle=0,scale=0.3]{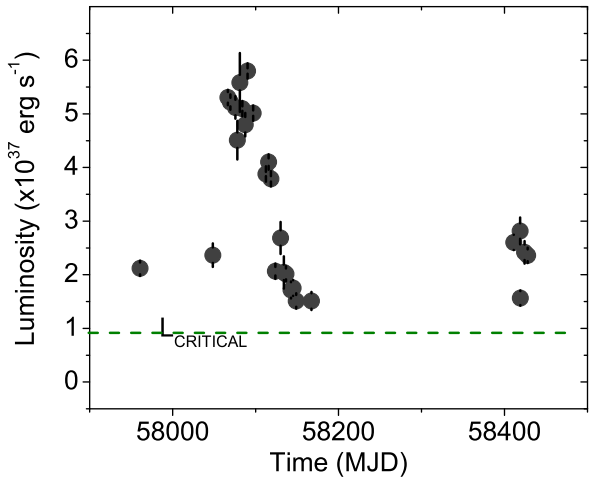}
\end{center}
\caption{Variation of luminosity with time in the energy range (1-10) keV for \textit{Swift/XRT}.} 
\label{8}
\end{figure}

\begin{figure}

\begin{center}
\includegraphics[angle=0,scale=0.3]{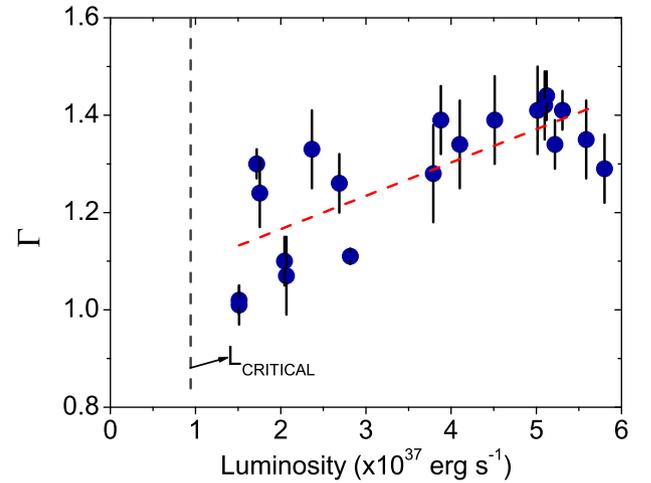}
\end{center}
\caption{Variation of photon indices ($\Gamma$) with the source luminosity in the energy range (1-10) keV observed by \textit{Swift/XRT}.}
\label{9}
\end{figure}

The spectral analysis of the \textit{Swift/XRT} data in the energy range (1-10) keV from (57960-58419) MJD covers the entire outburst of 2017 as well as few observations more than 300 days after the peak outbursts as shown in Table 3. The observations were carried out in both Windowed Timing Mode (OB IDs: 00010210002-00010210023) and in Photon Counting Mode (OB IDs: 00088639002-00088639003, 00010210034-00010210037) for analysis. The source spectrum in the soft energy range was fitted with an absorbed powerlaw. The hydrogen column density was fitted using TBabs component with $n_{H}$ ranging from 0.33$\pm$0.07--1.01$\pm$0.2 $cm^{-2}$. We could not constrain the $n_{H}$ parameter for few observations. Therefore, we fixed their respective $n_{H}$ to the galactic value $0.81\;\times\;10^{22}\;cm^{-2}$. The flux of the system varied over the course of the outburst, reaching a maximum value of $\sim1.26\;\times\;10^{-10}\;erg\;cm^{-2}\;s^{-1}$ at peak outburst and a minimum value of $\sim7.97\;\times\;10^{-12}\;erg\;cm^{-2}\;s^{-1}$ post outburst. The source luminosity varied between $\sim(10^{36}-10^{37})$ reaching its maximum at $\sim\;5.80\times10^{37}\;erg\;s^{-1}$ and minimum value at $\sim\;3.66\times10^{36}\;erg\;s^{-1}$ assuming a distance of 62 kpc \citep{Graczyk2014} as shown in Figure \ref{8}. Interestingly, it was observed that the source photon indices correlates with the source luminosity (see Figure \ref{9}).  

\subsection{Phase-Resolved Spectroscopy}
\begin{figure*}

\centering
\includegraphics[angle=0,scale=0.5]{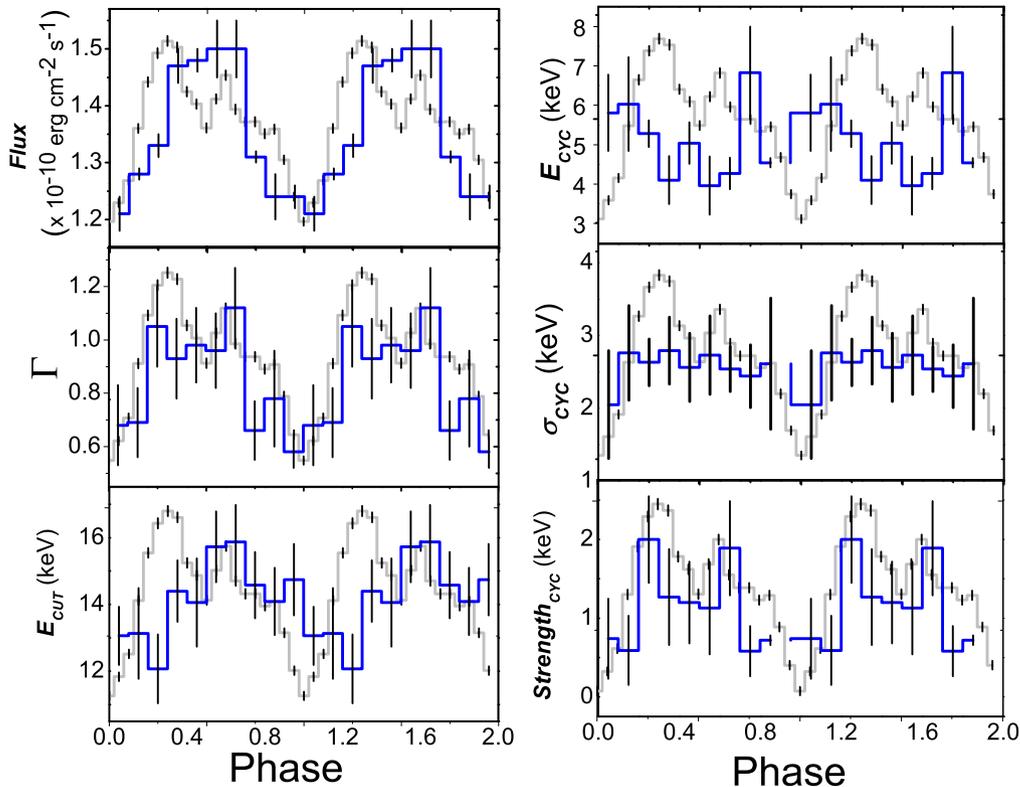}

\caption{Variation of spectral parameters with pulse phase. $\Gamma$ and $E_{CUT}$ represents Photon Index  and cutoff energy of cutoffpl model and \emph{$E_{CYC}$}, $\sigma_{CYC}$ and $Strength_{CYC}$ represents the Cyclotron line energy, width and the strength of the line corresponding to gabs component. Flux were calculated within energy range (3-79) keV for the \textit{NuSTAR} observation. The variation shown in the background in grey colour represents the (3-79) keV pulse profile. Errors quoted for each parameter are within 90\% confidence interval. }
\label{10}
\end{figure*}

In this section, we explore the evolution of spectral parameters with respect to the pulsed-phase. For this, we divided the spin period of the source SXP 15.3 into 10 segments of equal width. Next, we created good time interval (\texttt{gti}) files for each segment using \textit{task} \texttt{xselect} and merged them with \textit{task} \texttt{mgtime}. Using gti files merged by \texttt{mgtime}, we ran \texttt{nuproducts} for each segment to get the required spectra files corresponding to 10 phases.

The (3-79) keV spectra  corresponding to all the 10 segments were modeled with the  \textsc{model} combination constant$\times$TBabs$\times$(cutoffpl+gaussian)$\times$gabs applied for average spectra. The hydrogen column density ($n_{H}$) of TBabs component was fixed to the Galactic value for all the ten phase intervals. The flux in the (3-79) keV energy range varied between (1.21$\pm$0.03 to 1.50$\pm$0.06)$\;\times\;10^{-10}\;erg\;cm^{-2}\;s^{-1}$ and follows the main pulse profile as evident from Figure \ref{10}. The Photon Index ($\Gamma$) was seen to be correlated with the flux having maximum and minimum values between (1.12$\pm$0.09 to 0.58$\pm$0.06). The cutoff energy ($E_{c}$) parameter of the cutoffpl model was seen to vary at all phases having maximum value in the phase interval (0.6--0.7) and minimum values in the phase interval (0--0.1). The cyclotron line parameters were present and varied at all phases except in the last phase interval. The cyclotron line energy ($E_{CYC}$) varies between 25 \% to 11 \% about the mean value $\sim$ 4.8 keV having a maximum value of $\sim$ 6 keV in the phase interval 0.1-0.2 and a minimum value of $\sim$ 4.27 keV in the phase interval 0.6-0.7. The peak value of 6 keV is, however, associated with a higher uncertainty in its measurement. The width ($\sigma_{CYC}$) and strength ($Strength_{cyc}$) of the cyclotron line energy varies between (2.05$\pm$0.74 to 2.79$\pm$0.46) keV and (0.74$\pm$0.01 to 2$\pm$0.03) keV.

\section{Discussion and Conclusion}
\begin{figure}

\begin{center}
\includegraphics[angle=0,scale=0.3]{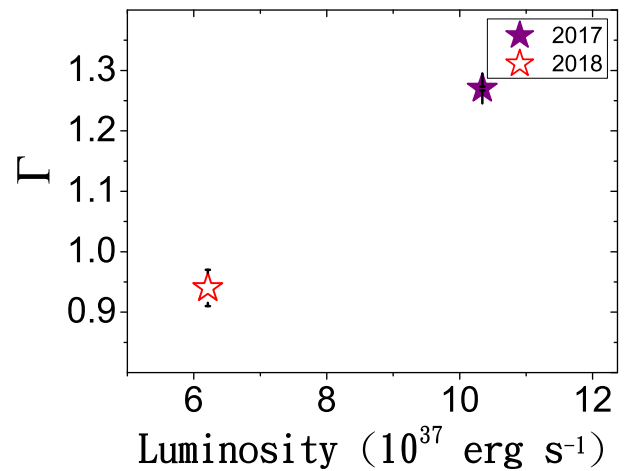}
\end{center}
\caption{Variation of Photon Index ($\Gamma$) with luminosity in the (3-79) keV energy range observed by \textit{NuSTAR}. }
\label{11}
\end{figure}
The temporal and spectral features of the BeXRB source SXP 15.3 have been analyzed in detail using the \textit{NuSTAR} observations of 2017 and 2018. Through timing analysis, we estimated the pulsation of the source (OB ID: 30361002004) at 15.2388\;$\pm$\;0.0002 s. The same analysis of \textit{NuSTAR} observation (OB ID: 30361002002), a year ago, detected the pulsation at 15.2534 $\pm$ 0.0004 s. As these two observations are taken at different spans of time, it appears that the source has spun up at a rate of --0.0176$\pm$0.0002 s $yr^{-1}$.

The shape of the pulse profile in our timing analysis evolves with energy and time. In the lower energy band (3-4) keV, the profile is characterized by a single asymmetric peak. However, in the lower intermediate energy bands ((4-5) keV and (5-7) keV, the profile seems to be marking weak dual peaked nature which is generally suggestive that the source is accreting in the supercritical regime. Above 7 keV, the profile loses its dual-peaked nature characteristics and at energy above 16 keV, the profile seems to be complex with a broad single peak of saw-tooth shape. The overall nature of the pulse profile is not sinusoidal in nature which indicates the pattern of harmonics as observed in the power spectrum of the source (see Figure \ref{1}). As evident from Figure \ref{4}, the pulse profile have evolved with time thereby indicating some activities in the accretion process around the source.

The PF of the source in the soft energy state shows decreasing trend up to $\sim$ 14 keV and then increases monotonically at higher energies. The latter one \textit{i.e} increase of PF with energy is typical for most X-ray pulsars \citep{Lutovinov_Tsygankov2009}. The fall in the PF around (12-16) keV looks quite intriguing. Temporal analysis of the source can also provide an indirect hint about the presence of CRSF. The decrease in the PF around the cyclotron line energy has been observed in the past for some X-ray pulsars  \citep{Tsygankov2007,Ferrigno2009,Lutovinov_Tsygankov2009,Tsygankov2010,Lutovinov2017} As evident from Figure \ref{5} during the 2017 outburst, one can see a fall in the PF around the cyclotron line energy range (5-10) keV. The search for CRSF feature in the spectra yielded a 5 keV cyclotron line \citep{Maitra2018}. However, when the flux of the source was 40\% less than the 2017 outburst, the decrement in the PF was observed in the same energy range in addition to decrement in the (12-16) keV energy range. The variation of pulse fraction with energy motivated us to check out the presence of CRSF feature or its harmonics (if any). However, no other harmonics were present other than the 5 keV cyclotron line both in the average spectra and phase resolved spectra.

The \textit{NuSTAR} spectra were fitted with different continuum models that include cutoffpl, comptt, and highecut. The average flux of the source in the energy range (3-79) keV is $\sim3.7\;\times\;10^{-10}\;erg\;cm^{-2}\;s^{-1}$ and its corresponding luminosity is $6\times\;10^{37}\;erg\;s^{-1}$ assuming a distance of 62 kpc. In order to verify the previous observation of Cyclotron line at 5 keV in the \textit{NuSTAR} spectrum \citep{Maitra2018}, the spectra of the source requires an additional absorption component at $\sim$ 5 keV in the (3-79) keV energy band. This suggests that the magnetic field of the source is $\sim$4 $\times\;10^{11}\;G$ using the expression $B=\frac{E_{CYC}(1+z)}{11.57} 10^{12}\;G$, where $E_{CYC}$ is the centroid energy of the CRSF in units of keV and $z$ is the Gravitational redshift parameter \citep{Harding_Daugherty1991}.

The change in the beaming pattern from pencil beam to fan-beam occurs at a luminosity referred to as Critical Luminosity ($L_{c}$) \citep{Becker2012}. A pencil-beamed X-ray emission is produced when accreting material directly falls on the surface of a neutron star if the luminosity of a pulsar is below $L_{c}$ (subcritical regime). With regard to the "pencil beam" pattern, the accreted material reaches the neutron star surface either by coulombic collisions with thermal electrons or nuclear collisions with atmospheric protons \citep{Harding1994}. X-ray emission escapes from the column's top in the pencil beam pattern \citep{Burnard1991}. A fan-beam pattern is characterized by luminosities greater than the $L_{c}$. In the case of supercritical regime, the material flow slows down towards the neutron star surface due to a radiation-dominated shock created in the accretion column. The presence of high radiation pressure is enough to stop the accreting matter above the neutron star. As a result, the shape of X-ray emission takes place in the form of a fan-beam pattern from the sides of the accretion column \citep{Davidson1973}. Furthermore, as the accretion rate increases, the shock region shifts higher in the accretion column, where the strength of the magnetic field is weaker. However, the beaming pattern now can be more complex as compared to a simple fan or pencil beam \citep{Kraus1995,Becker2012,Blum_Kraus2000}. In the  present work observed by \textit{NuSTAR}, the estimated value of luminosity is $6\times\;10^{37}\;erg\;s^{-1}$ for this source which is well above the critical luminosity of $\sim 9\;\times\;10^{36}\;erg\;s^{-1}$ indicating that the source accretion in the supercritical regime. The double-peaked feature of pulse profile in our timing analysis further supports a fan beam pattern in the NS radiation. 
%and the variation of photon index with the luminosity as shown in Figures 9 and 11
The long scale evolution of luminosity with time as shown in Figure \ref{8} indicates the source accretion in the super-critical regime during its major outburst in 2017 observed by \textit{Swift/XRT}. Post outburst, at $\sim$ 58411-58430 MJD, which corresponds to the observations very close to the \textit{NuSTAR} data, the source is clearly accreting in the super-critical regime.

A negative correlation between the photon index and the source luminosity has been observed earlier in the sub-critical regime \citep{Muller2013}. Sources that showed negative correlation with luminosity earlier in the subcritical regime are 1A 0535+262, 1A 1118–612, GRO J1008–57, XTE J0658–073 while sources that underwent transition are 4U 0115+63, EXO 2030+375, and KS 1947+300 \citep{Reig2013}. The photon index of the source showed correlation (see Figure \ref{9} and \ref{11}) with increasing luminosity implying the source accretion to be in supercritical-critical regime above the critical limit. In addition, pulse phase resolved spectroscopy shows that the photon index becomes softer when the source flux is high and when the source flux is low, the photon index becomes harder, thereby, indicating the source to be in the supercritical regime at $\sim$ 58418.86 MJD. The variation shown in phase-resolved spectral analysis further supports the results.

The CRSF feature are more prominent in the phase-resolved spectra and are known to demonstrate variations with the pulse phase \citep{Burderi2000,Kreykenbohm2004}. Pulse phase revolved analysis of CRSF feature allows us to understand the emission geometry and to map the magnetic field structures at different viewing angles \citep{Maitra2013}.
% Therefore, considering such possibility, we checked for phase transient absorption line in the phase-resolved spectra, but did not find any positive indication in favour of cyclotron absorption lines.  
The pulse phase-resolved spectral analysis as shown in Figure \ref{10} shows that the spectral parameters of different components vary with pulse phases. Indeed, it is obvious from Figure \ref{10} that the cyclotron line energy ($E_{CYC}$) varies at all phases except near off-pulse phase which is consistent with the works of \cite{Maitra2018}. The numerical simulations of the cyclotron line indicate that the line parameters are found to vary roughly by 10–20 \% with the pulse phase \citep{Araya_Gochez_Harding,Schonherr2007,Mukherjee2012}. These cyclotron line parameters can change over the course of a pulse phases by more than 20 \% and as much as 40 \% \citep{Staubert2014}. For instance, the centroid energy variation for Her X-1 (the first discovered CRSF) is of the order of 25 \% \citep{Staubert2014}. The lower value of cyclotron line energy with phase is $\sim$4.27 keV, 11 \% lower than the average value. Therefore, both the peak value and the lower value of the cyclotron line variation with phases are within the upper limit and the lower limit discussed in the section.  

%It is noted that the line's strength and width are incredibly low when compared to other phases. One reason for such a low strength may be due to onset of partial eclipse of the accretion column by a neutron star at certain phases. The other reasons can be due to the large gradient in the magnetic field's strength over the visible column height or latitudes on the surface of the neutron star that may lead to very low strength and width of the line at certain phases. 

In summary, we present a detailed timing and spectral results of SXP 15.3 observed by \textit{NuSTAR} and \textit{Swift/XRT} observations. The presence of CRSF at $\sim$ 5 keV is clearly noticeable and shows variability with pulse phase resolved spectroscopy. The source pulse period between the two \textit{NuSTAR} observation indicates that the source has spun up at a rate of --0.0176 s $yr^{-1}$. \textit{Swift/XRT} observation on a long time scale shows that the source luminosity lies above Critical luminosity therby indicating the source accretion to be in super-critical regime. The result is further supported by the spectral softening with the increase in luminosity. 

 \section*{Data availability}
 
 In this reasearch work, we have used the publicly available observational data which can be accessed from the HEASARC data archive.

\section{Acknowledgement}
This research work has made use of the publicly available \textit{NuSTAR} data of the pulsar provided by the NASA High Energy Astrophysics Science Archive Research Center (\texttt{HEASARC}) online service maintained by the Goddard Space Flight Center. The \textit{NuSTAR} Data Analysis Software (\texttt{NuSTARDAS}) jointly developed by the ASI Space Science Data Center (SSDC, Italy) and the California Institute of Technology (Caltech, USA) has been used for our study. We would like to thank the anonymous reviewer for his/her suggestions and comments that helped us in improving the manuscript in the current version. Authors are thankful to the IUCAA Center for Astronomy Research and Development (\texttt{ICARD}), Physics Department, NBU for extending research facilities.

%%%%%%%%%%%%%%%%%%%%%%%%%%%%%%%%%%%%%%%%%%%%%%%%%%



%\bibliographystyle{mnras}


%%%%%%%%%%%%%%%%%%%%%%%%%%%%%%%%%%%%%%%%%%%%%%%%%%

%%%%%%%%%%%%%%%%% APPENDICES %%%%%%%%%%%%%%%%%%%%%

%%%%%%%%%%%%%%%%%%%%%%%%%%%%%%%%%%%%%%%%%%%%%%%%%%

% Don't change these lines
\bsp	% typesetting comment
\label{lastpage}
\end{document}